\documentclass[galaxies,article,accept,moreauthors,pdftex]{Definitions/mdpi} 

\firstpage{1} 
\pubvolume{10}
\issuenum{1}
\articlenumber{37}
\pubyear{2022}
\copyrightyear{2022}
\externaleditor{Academic Editor: Martín Guerrero, Noam Soker and Quentin A. Parker 
} 
\datereceived{20 January 2022} 
\dateaccepted{15 February 2022 } 
\datepublished{17 February 2022} 
\hreflink{https://doi.org/10.3390/\linebreak{galaxies10010037}} 

\Title{Ionization-Gasdynamic Simulations of Wind-Blown Nebulae around Massive Stars}

\TitleCitation{Ionization-Gasdynamic Simulations of Wind-Blown Nebulae around Massive Stars}

\Author{Vikram V. Dwarkadas 
 \orcidA{}}

\AuthorNames{Vikram V. Dwarkadas}

\AuthorCitation{Dwarkadas, V.V.}

\address[1]{%
Department of Astronomy and Astrophysics, University of Chicago, 5640 S. Ellis Ave., ERC 569, Chicago, IL 60637, USA; vikram@astro.uchicago.edu; Tel.: +1-773-834-4668}

\abstract{Using a code that employs a self-consistent method for computing the effects of photo-ionization on circumstellar gas dynamics, we model the formation of wind-driven nebulae around massive stars.  We take into account changes in stellar properties and mass-loss over the star's evolution.  Our simulations show how various properties, such as the density and ionization fraction, change throughout the evolution of the star. The {multi-dimensional simulations} reveal the presence of strong ionization front instabilities in the main-sequence phase, similar to those seen in galactic ionization fronts. Hydrodynamic instabilities at the {interfaces} lead to the formation of filaments and clumps that are continually being stripped off and mixed with the low density interior. Even though the winds start out as completely radial, the spherical symmetry is quickly destroyed, and the shocked wind region is manifestly asymmetrical. The simulations demonstrate that it is important to include the  effects of the photoionizing photons from the star, and simulations that do not include this may fail to reproduce the observed density profile and ionization structure of wind-blown bubbles around massive stars. }


\newcommand{\beqn}{\begin{eqnarray}}
\newcommand{\eeqn}{\end{eqnarray}}
\newcommand{\be}{\begin{equation}}
\newcommand{\ee}{\end{equation}}

\newcommand{\msun}{\mbox{$M_{\odot}$}}

\newcommand{\bfig}{\begin{figure}[h]}
\newcommand{\efig}{\end{figure}}

\begin{document}

\section{Introduction}

\label{sec:intro}
Massive stars ($\geq$10$\msun$) lose mass throughout their lifetime, via winds and eruptions. Many will subsequently end their lives in a cataclysmic supernova (SN) explosion, while others may collapse directly to a black hole \citep{sukhboldetal16}. The interaction of the wind material with the surrounding medium, {which could be the interstellar medium (ISM)} or a wind from a prior phase, creates vast wind-blown cavities surrounded by a dense shell, referred to as wind-blown bubbles. Simultaneously,
radiation from the star can ionize the surrounding medium. As the star
evolves through various stages, the mass-loss parameters, and the
ionizing flux or number of ionizing photons, will change, affecting
the structure of the bubble. If the star ends its life in a SN explosion, the
resulting SN shock wave will expand within the bubble, and the
dynamics and kinematics of the shock wave, dependent on the bubble
parameters, will differ from evolution within the ISM \cite{dwarkadas05}. SN 1987A was a well observed case of a SN exploding in the wind bubble created by the progenitor star, where the observed emission was clearly affected by the structure of the bubble \citep{cd95}. The relativistic blast wave associated with a gamma-ray burst (GRB) is also expected to expand within wind bubbles surrounding Wolf-Rayet (W-R) stars, which will affect the radiation signatures \citep{wb06}. It is
therefore important to understand the precise structure of the bubble, as it influences the evolution of the subsequent SN shock wave, and the resulting emission from the SN.

The structure of wind bubbles was initially elucidated by Weaver et al.~\cite{weaveretal77}. Proceeding radially outward from the star, they delineated 4 regions: (1) a freely expanding wind region, (2) a shocked wind region, (3) the shocked ambient medium, forming a thin dense shell, and (4) the unshocked ambient medium. An inner, wind termination shock separates region (1) and (2), a contact discontinuity {separates} regions (2) and (3), and an outer, generally radiative shock {separates} regions (3) and (4). Since the shock is radiative, the material behind it tends to pile up {into} a thin shell of high density. The radius of the wind bubble $R_b$ depends on the wind mass-loss rate and velocity and the density of the surrounding medium, and increases with time as $R_b \propto t^{3/5}$, if the bubble is expanding into a constant density surrounding medium. 

Since the seminal work of \cite{weaveretal77}, the hydrodynamics of wind bubbles around massive stars has been explored by many authors. Initial models considered the general interaction of a fast wind with the interstellar medium, or with a slower wind from the previous epoch  \cite{rozyczka85,bd97}.  Others modelled the morphology of asymmetrical bubbles, especially around Luminous Blue Variable stars \cite{bl93,db98b, frd98, do02}. Garcia-Segura et al. \cite{glm96, gml96} explored realistic models of wind evolution, where the wind parameters were obtained directly from stellar evolution calculations. They calculated models for the wind bubble around a 35 $\msun$ star and a 60 $\msun$ star. Since then many other authors have explored the evolution of the bubbles using analytical calculations as well as multi-dimensional simulations, including  \citet{dwarkadas05}, Freyer et al. \cite{fhy03}, van Marle et al. \cite{vlg05,vlag06}, Freyer et al. \cite{fhy06}, \citet{dwarkadas07}, \mbox{Chita et al. \cite{chitaetal08}}, \citet{vk12}. 

In order to model the structure of wind-bubbles accurately, the models need to take into account the combined influence of the stellar wind as well as the ionizing radiation from the central star. Analytic calculations including a  photoionized wind were made by \cite{chevalier97}. Early numerical models that explored the evolution of the medium surrounding massive stars \cite{glm96, gml96} dealt with ionization using a Stromgren sphere approximation. These authors did not model the main-sequence (MS) stage in multi-dimensions. Similar limitations were included in the models of \cite{vlg05, vlag06}. A better treatment of ionization properties was included in \cite{fhy03, fhy06}. \citet{dwarkadas07} considered the entire evolution in multi-dimensions, and \citet{dwarkadas08} studied the turbulence within the interior in detail, but  the effects of stellar ionization were not included in these calculations. 3D simulations carried out by \cite{vk12} also did not include the effect of the ionizing photons. Models that included a comprehensive treatment of both the ionization and the gasdynamics were published by \mbox{\citet{ta11}}, followed by \citet{dr13}. HII regions around moving O stars were modelled by Mackey et al. \cite{mlg13}, {while superwinds from a cluster of stars were examined by Danehkar et al. \cite{dog21}. Three-dimensional hydrodynamic simulations of efficiently cooled stellar wind bubbles, without including the effects of photoionization, were carried out by Lancaster
et al. \cite{lancasteretal21}.}  A series of papers highlighting different aspects of Wolf-Rayet nebulae \cite{mpp20, meyeretal20,meyer21} also did not include the effects of the ionizing radiation from the star. 

An enduring question in the study of wind blown bubbles is that of the temperature of the shocked wind bubble region. The wind velocities of massive stars in the main-sequence and Wolf-Rayet stages are 1000--3000 km s$^{-1}$. The temperature of the shocked wind region beyond the wind termination shock, which makes up most of the volume of the bubble, would therefore be expected to exceed 10$^7$ K. In terms of X-ray emission, the X-ray temperature should be greater than about 1 keV. The hot bubbles should be visible as regions of diffuse X-ray emission. However, despite extensive searches with Chandra and XMM, diffuse X-ray emission has been detected in only few cases \cite{cgg03,wriggeetal05}, with observed X-ray luminosities lower than expected. The inferred X-ray temperature for NGC 6888 and S308 is a few times 10$^6$ K, also lower than expected. \citet{zp11} find a higher temperature component ($>$2 keV) from {\it Suzaku} data of NGC 6888, but this is not supported by subsequent {\it Chandra} analysis \cite{toalaetal14, toalaetal16}. Toalá et al. \cite{toalaetal15} find a high temperature component $>$$4.5$ keV for NGC 2359, although it contributes to less than 10\% of the total X-ray flux. In most other cases, diffuse X-ray emission is not detected at all. This suggests the existence of some mechanism that tends to lower the interior temperature, or equivalently that the energy that would go into raising the temperature is being expended elsewhere. 

\emph{The  goal of this project is to carry out ionization-gasdynamic simulations of wind-blown bubbles followed by subsequent X-ray modelling. We will assess accurately the interior structure of wind bubbles by studying the evolution in multi-dimensions, and use the results to compute the X-ray temperatures and examine the energy budget.} We will use a code that combines the effects of photoionization from the star with the gasdynamics. As shown in \cite{ta11,dr13}, the ionizing photons from the star can modify the hydrodynamics, and thereby the density structure of the bubble, which can alter the subsequent emission from the system. Therefore it is essential that the ionizing radiation from the central star be taken into account. In this paper we elaborate significantly on the work first outlined \mbox{in \citet{dr13}}, describing in detail the gasdynamic simulations, the density structure, as well as the onset and growth of instabilities in models of ionized wind blown bubbles around massive stars. We use a code, AVATAR, introduced in \citep{dr13}, to carry out moderate resolution simulations of wind blown bubbles around massive stars.  Our simulations improve on most previous work in that we model all the phases of the star's evolution, starting from the main-sequence, in multi-dimensions. {This} was not the case in much of the {earlier} modelling. {Our simulations} reproduce hydrodynamic instabilities as well as those due to the ionizing radiation. The  results of our modelling work are consistent, and often better, than previous models that studied wind-bubbles around massive stars. The simulations are used to compute the X-ray emission from the bubbles. The X-ray computation methodology and results will be described in detail in a follow-up paper. 

This paper proceeds as follows: In Section \ref{sec2} we briefly outline the photoionization code. In Section \ref{sec3} we use the code to carry out spherically symmetric simulations of wind-blown bubbles around a 40 $\msun$ star. Section \ref{sec4} shows the results for wind bubbles around a 40 $\msun$ star computed in 2-dimensions. Summary and conclusions are given in Section \ref{sec5}.

\section{AVATAR Code}\label{sec2}

An initial version of the code used in our modelling effort was described in \cite{rosenberg95}. Substantial modifications were made to enable the current modelling effort. AVATAR contains a self-consistent method for computing the effects of photo-ionization on circumstellar gas dynamics. The effects of geometrical dilution and column absorption of
radiation are considered. The gasdynamic algorithm makes use of a multidimensional covariant implementation of well established Eulerian finite difference algorithms. A second-order van Leer monotonic transport algorithm is used for the advection of total mass and the neutral component, and a third order piecewise parabolic algorithm scheme is available as needed.  Tabulated functions are used to compute the collisional ionization rate and cooling function. Shocks are treated using an artificial viscosity. Grid expansion, very useful when studying systems which expand over several orders of magnitude over the course of the simulation, is incorporated. The method operator splits the contribution due to photoionization effects from the usual gas dynamics, and utilizes a backward-Euler scheme together with a Newton-Raphson iteration procedure for achieving a solution. The algorithm incorporates a simplified model of the photo-ionization source, computes the fractional ionization of hydrogen due to the photo-ionizing flux and recombination, and determines self-consistently the energy balance due to ionization, photo-heating and radiative cooling. Our method is comparable to that used by \cite{ta11}, and superior to that of other calculations that use the on-the-spot approximation, and do not take recombination into account \citep{glm96, gml96, vlg05, vlag06}.

\section{Spherically Symmetric Simulations}\label{sec3}

\begin{figure}[H]
\begin{adjustwidth}{-\extralength}{0cm}
\centering
\includegraphics[width=\columnwidth]{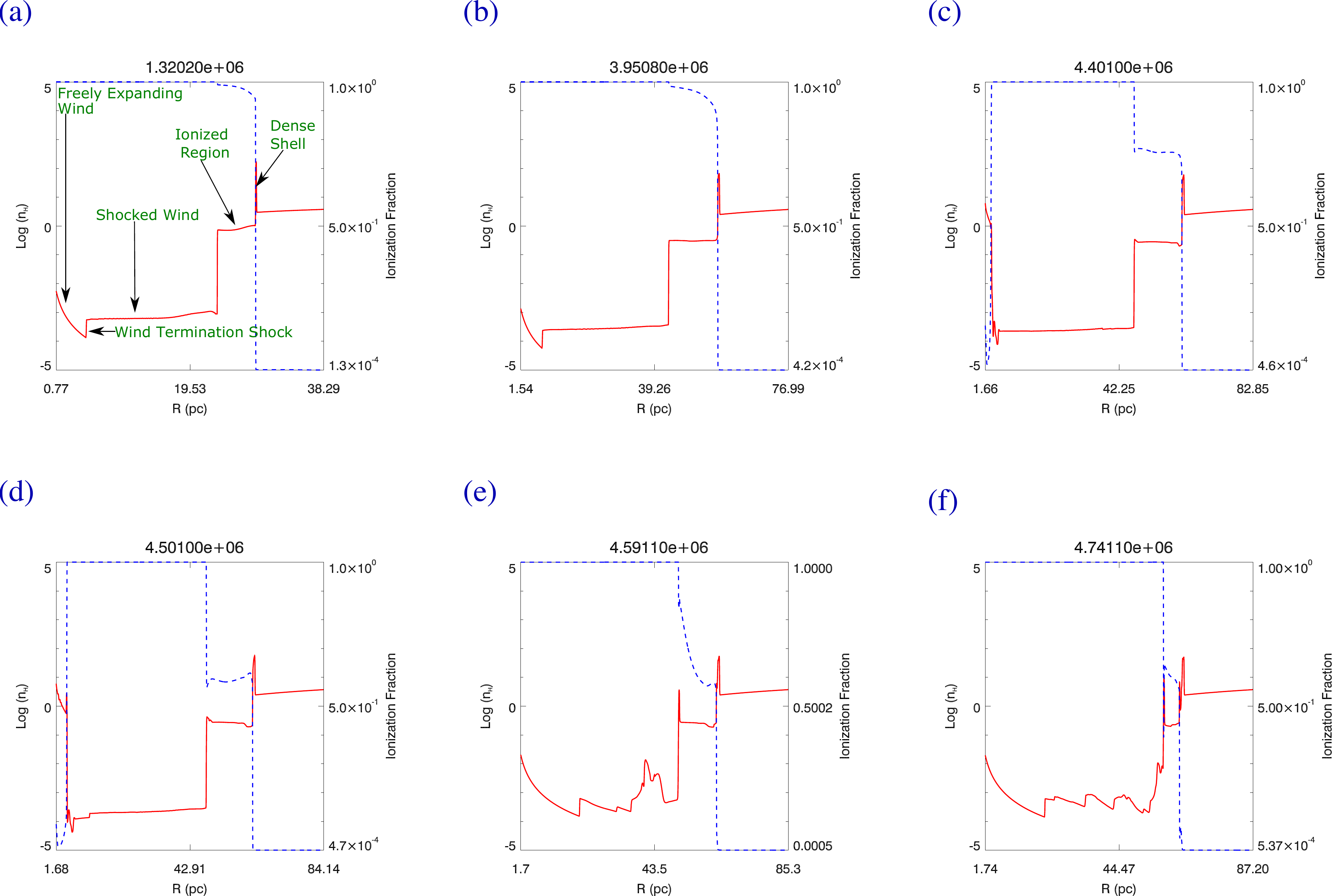}
\end{adjustwidth}
\caption{ The number density (red line) and ionization fraction (blue line) at various times, from the spherically symmetric simulation. The left scale in each panel is the logarithm of the number density $n_H$, defined as $n_H=\rho/({1.67} \times 10^{-24}$) where $\rho$ is the gas density. The right hand side is the ionization fraction, stretching from a  value close to zero (almost neutral) to 1 (fully ionized). The stellar evolution time (age of the star) in years is given at the top. Panels (\textbf{a},\textbf{b}) refer to the MS phase, panels (\textbf{c},\textbf{d}) to the RSG phase, and panel (\textbf{e},\textbf{f}) to the W-R phase. \label{fig:den1d}}
\end{figure}

In order to illustrate the salient features of the density and ionization structure of the wind-bubble bubble during the various phases of stellar evolution, we first consider spherically symmetric simulations of the wind bubble around a 40 $\msun$ star.  The simulation shown, carried out with the AVATAR code, is run in spherical ($r, \theta$) co-ordinates, with 2400 radial zones. The star starts off as a massive main-sequence (MS) star, moving off the MS sequence to become a red supergiant (RSG) after about 4.3 million years. After another 200,000 years in the RSG phase, it loses its outer layers to become a Wolf-Rayet (W-R) star.  The wind and ionization parameters for the star in each phase, listed in Table \ref{tab:bub}, are taken from \cite{vlg05}, who obtained them from the stellar evolution calculation of \cite{ssmm92}. During each phase, the wind and ionization parameters are assumed constant, following \cite{vlg05}, although in reality they would be expected to vary with time. The initial setup consists of a freely expanding wind with the mass-loss rate and wind velocity of the star in the MS stage, running into a constant density external medium. The initial grid extends from 0.25 to 12.5 pc. As the bubble expands and the outer shock moves outwards, the grid expands along with it. The number density and ionization fraction at various times in the evolution is shown in Figure \ref{fig:den1d}. The increasing grid size at each step is reflected in the different panels in Figure \ref{fig:den1d}. The simulation shown ran for 1,336,076 timesteps.

\begin{table}[H] 
\small
\caption{The values  {of relevant} parameters in the various phases of the evolution of a 40 $\msun$ star, taken from \cite{vlg05}.\label{tab:bub}}
\newcolumntype{C}{>{\centering\arraybackslash}X}
\begin{tabularx}{\textwidth}{lCCC}
\toprule
\textbf{Parameter}	& \textbf{MS}	& \textbf{RSG} & \textbf{W-R}\\
\midrule
Mass-loss Rate ($\msun$yr$^{-1}$)	& 9.1 $\times\ 10^{-7}$	& 8.3 $\times\ 10^{-5}$ & 4.1 $\times\ 10^{-5}$ \\
Wind Velocity	(km s$^{-1}$)	& 890			& 15 & 2160 \\
{Star age at} End of Phase ($\times\ 10^6$ yr) & 4.309   &  4.508 & 4.786 \\
Ionizing Photons ({s$^{-1}$}) &  4.62 $\times\ 10^{47}$ & 3. $\times\ 10^{41}$ & 3.86 $\times\ 10^{47}$ \\
\bottomrule
\end{tabularx}
\end{table}

In about 100 timesteps after the start of the simulation, the density structure begins to resemble the wind-blown structure depicted in \cite{weaveretal77}. It takes a few hundred more steps before the density structure attains the self-similar shape shown in Figure \ref{fig:den1d}a. Going outwards in radius, we identify the freely expanding wind, wind termination shock, and shocked wind region.  At this point however the structure deviates from the model of Weaver et al. \cite{weaveretal77}, in that we see a higher density ionized region, formed by the ionizing photons from the star ionizing the surrounding material, in this case the interstellar medium. The size of any such region depends on the star's surface temperature and number of ionizing photons from the star.  An ionized region such as this was identified in the structure of the wind bubble surrounding SN 1987A, in order to explain the increasing X-ray and radio \mbox{emission \cite{cd95}}. The ionized HII region ends in a contact discontinuity, outside of which lies the dense shell of swept-up surrounding medium, bounded on the outside by a radiative shock. This structure is consistent with that shown in \cite{ta11}. We note that although in this paper we continually refer to the outer higher density ionized region as the HII region, in principle the shocked wind bubble is also ionized by the star and would be considered a part of this HII region. We refer to them as separate entities to distinguish the wind material from the shocked ambient medium material. As seen in Figure \ref{fig:den1d}a, the ionization fraction throughout the bubble is close to 1, i.e. the bubble is essentially fully ionized in the MS phase. The ionization fraction drops to zero within the dense shell, showing that the ionization front is captured in the shell. Thus the shell itself remains partly ionized and partly neutral. The density of the shocked wind region is low, and its pressure (not shown) is quite high, therefore the temperature in this region can be high in spherically symmetric simulations.  Although the bubble expands in size, the structure is more or less self-similar, as can be seen in Figure
 \ref{fig:den1d}b, where the density and ionization structure at 3.95 million years closely resemble that at 1.32 million years (Figure \ref{fig:den1d}a), even though the bubble size has increased by a factor of 2. 

At around 4.3 million years, the star leaves the MS and forms a RSG star. The velocity of the RSG wind is only 15 km s$^{-1}$, much lower than that of the MS wind, but its mass-loss rate is significantly higher. Therefore a new pressure equilibrium is established, though there may not always be enough time in the RSG phase for complete pressure equilibrium to be established. The high density RSG wind begins to expand within the MS wind.  The surface temperature of the RSG star is much lower than that of the MS star, and the number of ionizing photons considerably reduced. This has two effects - the RSG wind itself has a low ionization fraction, and the material in the outer HII region begins to recombine, since the star can no longer continue to ionize it. Therefore the ionization fraction in the HII region begins to decrease (Figure \ref{fig:den1d}c), and continues to do so during the RSG phase (\mbox{Figure \ref{fig:den1d}d}). The RSG wind begins to pile up against the MS wind. Although it is not clear from Figure \ref{fig:den1d}, the ionization fraction in the ionized HII region can fall to very low levels by the end of the RSG phase, depending on the length of the phase and the temperature of the star. Much of the mass loss from the star happens in the RSG phase, followed by the subsequent W-R phase.

After about 200,000 years in the RSG phase, the star begins to shed its outer layers, and forms a compact W-R star. In this phase, the stellar temperature, and the number of ionizing photons, is much higher, so the W-R star is able to slowly re-ionize the recombined material in the HII region (Figure \ref{fig:den1d}e,f). At the end of the simulation the bubble is once again almost fully ionized, which can be clearly seen in the 2D simulations below. The W-R wind has a velocity that is two orders of magnitude larger than the RSG wind velocity, and a mass-loss rate that is only a factor of 2 lower. The wind momentum therefore significantly exceeds that of the RSG wind. The result is that the RSG wind is pushed outwards by the W-R wind, which mixes the RSG material with that of the MS wind. At the end of the W-R phase, the resultant density structure, although more complicated, in essence resembles that of the Weaver et al. \cite{weaveretal77} model, albeit with considerable density fluctuations. The {W-R} wind also pushes out against the HII region, which is compressed. In the 1D simulation, the W-R wind is not able to fully push out all the way to the dense shell. In the 2D simulation (below), the wind is able to make its way all the way to the dense shell in some regions, and the extent of the ionized region is greatly reduced.

The resultant W-R nebula consists of a mixture of MS, RSG and W-R material. As the outer shock expands and constantly sweeps up material, which collects in the dense shell, the shell mass can be several hundred solar masses. The bubble radius is set essentially in the MS phase, as the later wind stages do not have sufficient momentum to push the shell much further out. In this simulation there is $<$$10\%$ increase in radius after the MS stage.  Most of the material in the interior of the bubble arises in the post-main-sequence phases, when the star loses much of its mass. The abundance of the material within the W-R bubble will reflect mass lost in all the prior phases. 

We note that the density structure of the bubble described here may deviate somewhat from that in papers that did not include both ionization and recombination. In many simulations, the effects of photo-ionization are not included at all, and there is no ionized HII region present. In simulations where ionization was included in terms of the Stromgren radius, such as in \citep{vlg05}, an ionized HII region is formed; however, in the RSG stage, the Stromgren radius reduces sharply due to the low surface temperature and lack of ionizing photons, and they {refer to it as} the `former' HII region, since  {in their method} recombination takes place instantly. In reality, the recombination process takes time (recombination time t$_{rec} \propto 3 \times 10^{12}/n_H$ s, \citep{osterbrock89}), and could require hundreds to thousands of years depending on the density ($n_H$) of the region, and the number of ionizing photons emitted by the star. The effects are more complicated than can be dealt with in the Stromgren approximation. The physical processes included in our code are most similar to those in \cite{ta11}, and it is no surprise that our results are compatible with theirs. Considering the different codes and methods used, the consistency of the results is heartening, and a confirmation of both~codes.

\section{Two Dimensional Simulations}\label{sec4}
\subsection{Hydrodynamics}

Once the basic  density, pressure, temperature and ionization structure of the wind bubble have been understood, we proceed to compute the evolution of the wind bubble in two dimensions (2D). This allows us to study the breaking of spherical symmetry, the onset of instabilities and the growth of turbulence within the bubble. {Our work differs from most past simulations in that we compute the evolution in 2D starting from the main sequence (MS) phase. Previous simulations by Garcia-Segura et al. \cite{glm96, gml96}, van Marle et al. \cite{vlg05}, and \citet{ta11} did not compute the MS stage in multi-dimensions, presumably due to the expense of the calculation.} Freyer et al. \cite{fhy06} did run 2D simulations of the evolution of a 35 $\msun$ star {starting} from the MS phase, and our results can be qualitatively compared to theirs, although variation in parameters, differences in the code and the grid setup, and the quality of their plots, precludes quantitative comparisons. We show, similar to the findings of Freyer et al. \cite{fhy06}, that the MS phase leads to substantial instabilities, the effects of which can persist throughout the evolution, and lead to increased turbulence within the wind blown region. 

We carried out a suite of runs for different mass stars using the (r,$\theta$) coordinate system, with an expanding grid containing 600 radial and 400 angular zones. The expanding grid ensures that the simulation is well resolved throughout most of the evolution. As in the 1D case, the results for a 40 $\msun$ star are presented here.

The initial density structure is similar to that in 1D, consisting of a freely expanding wind with the MS parameters interacting with a constant density ambient medium. Soon after the simulation is initialized, the basic density structure seen in the 1D simulations is reproduced, with a freely expanding wind, wind termination shock, shocked wind region (dark blue in figure), HII region, and dense swept-up shell (yellow in figure). The difference from the spherically symmetric simulation is that the regions are not spherically symmetric, but deviate from sphericity, and are subject to various types of instabilities.

\begin{figure}[H]
\includegraphics[width=\linewidth]{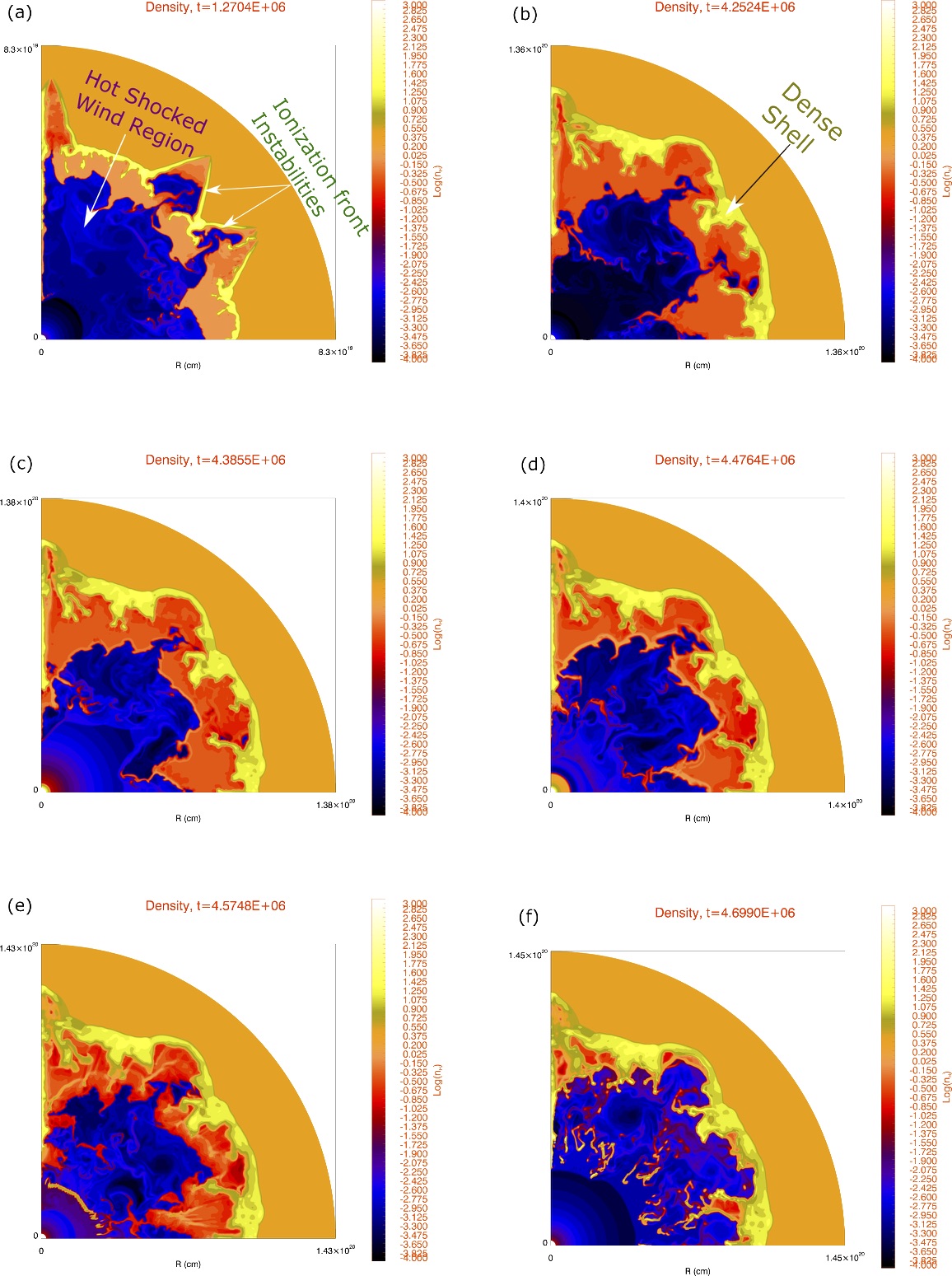}
\caption{ The density structure of the wind bubble from the 2D simulation. The scale in each panel is the logarithm of the number density $n_H$, defined as $n_H=\rho/({1.67} \times 10^{-24}$) where $\rho$ is the gas density. The stellar evolution time in years is given at the top. Similar to Figure \ref{fig:den1d}, panels (\textbf{a},\textbf{b}) refer to the MS phase, panels (\textbf{c},\textbf{d}) to the RSG phase, and panel (\textbf{e},\textbf{f}) to the W-R phase.  The hot shocked wind region (dark blue) becomes very asymmetric in shape as the simulation proceeds. The growth of ionization front instabilities in the MS phase, instabilities at the inner edge of the dense shell, and at the interface between the hot shocked wind region and the ionized HII region, is clearly seen, and is described in the text. }
\label{fig:den2d}
\end{figure}

Shock waves and contact discontinuities are known to be unstable to various types of dynamical instabilities, such as Kelvin-Helmholtz, Rayleigh-Taylor (or the related Richtmeyer-Meshkov) and the Vishniac-type instabilities \cite{vishniac83}. In the case of thin shells bound by radiative shocks, such as the dense swept-up shells in wind-bubbles, the latter instability can cause perturbations equal to or larger than the shell thickness to grow into finger-like projections, which may continue to grow non-linearly \citep[see for example][]{db98a}. Hydrodynamic instabilities in the dense shell of a wind bubble were highlighted by \citep{dwarkadas07}. These instabilities can be further exacerbated in the presence of an ionization front. We interpret the instabilities seen at the outer edge of the dense partly-ionized shell (Figure \ref{fig:den2d}a), leading to the growth of large perturbations, as an ionization-front instability associated with a D-type ionization front. Freyer et al. \cite{fhy03} note the existence of similar structures, which they suggest is due to an ionization front instability combined with the redistribution of mass (and opacity) by the stellar wind shell. In principle it is difficult to isolate these two effects as they influence each other. Although not visible in the figure, during the course of the evolution, photons can sometimes be seen to leak outside these large amplitude projections due to small gaps that may develop in the unstable structures. 

The  stability of an ionization-shock front, in the isothermal case without self-gravity, was examined by \citet{giuliani79}. Numerical simulations to study this instability were carried out by \citet{gsf96}, who discussed the evolution of ionization front instabilities in expanding HII regions. Similar instabilities are also seen in galactic ionization fronts, and 3D simulations carried out by \citet{wn08} show the presence of spike-like projections very similar to those seen in Figure \ref{fig:den2d}a (see their Figure \ref{fig:den2d}). The reason for the ionization instability growth \citep{wn08} is that when the surface is perturbed, ionizing photons tend to make their way out via regions of the lowest opacity,  generally regions of the lowest density, thus leading to transverse velocity gradients that accentuate the perturbations. As initially calculated by \citet{giuliani79}, shown in numerical simulations by \citet{gsf96}, and seen in our simulations, the wavelengths of the fastest growing modes tend to increase with time in the simulation.   It is clear that they create large amplitude, pronounced structures. The instabilities eventually die out in our simulations late in the MS phase (Figure \ref{fig:den2d}b), as the pressure imbalance tends to smooth out the velocity variations. The decreasing grid resolution may also play a role here;  as the grid expands in our simulations, the size of each grid cell in the outer regions is continually increasing, thus reducing the size of the perturbations compared to the grid size, and damping the hydrodynamic instabilities that initiate the process.

We also see finger-like projections of smaller amplitude at the inner edge of the dense shell. These are attributed to various instabilities, {combined with photo-evaporation} at the inner edge. The {HII region and shocked wind material together} push out on the decelerating higher density shell as the bubble expands, leading to `fingers' of higher density material expanding into the HII region.  {Velocity gradients at the inner edge lead to the formation of Kelvin-Helmholtz clumps}. These combine to form structures {reminiscent of} the photo-ionized pillars of emission seen in regions like the Eagle Nebula \citep{hesteretal96}. As the simulation proceeds, these {structures} are continually being broken off and mixed with the interior plasma, while more grow at the inner edge. This constant mixing of material leads to colder, denser material from the shell being continually mixed with the hot, lower density interior plasma.

The hot shocked wind region (dark blue in the figure) no longer remains spherically symmetric, as  seen in Figure \ref{fig:den2d}. In the simulations,  the HII region forms marginally {after} the hot shocked wind region forms. The interface separating the shocked wind region and the HII region is unstable. In the very initial stages, the shocked wind region is at a {slightly} higher pressure than the HII region. We then have a situation where a low density, high pressure region is pushing out against a higher density, low pressure region, and thus the interface is susceptible to the Rayleigh-Taylor instability. These instabilities perturb the interface, leading to velocity fluctuations within the bubble, which are accentuated by the turbulence in the hot shocked wind region. The velocity {gradients} can lead to clumps forming at the edges due to the Kelvin-Helmholtz instability. {As the simulation proceeds, we again see clumps and filaments continually breaking off and mixing with the low density shocked wind. The dense clumps may sometimes end up shielding the remaining column from the ionizing radiation.} At the same time the {unstable} dense swept-up shell leads to an asymmetry in the HII region shape. As the entire wind-blown structure expands, the pressure between the HII region and the shocked wind begins to equilibrate, but imbalance in pressure in localized regions further destroys the spherical symmetry of the shocked wind region.   In some regions the {shocked wind region} can push up all the way to the shell (Figure \ref{fig:den2d}a). 

Although not shown herein, the time variation of the radius of the inner wind termination shock results in a continuous deposition of vorticity within the hot shocked wind region, leading to the formation of large vortices in the shocked wind region, and thus significant non-radial velocities. This must be viewed with caution in a 2D simulation, because 2D turbulence leads to an inverse cascade in energy \cite{es04}, as compared to the direct cascade in 3D. We do expect some amount of similar turbulence in 3D however, since the variation of the wind termination shock position will still be present in 3D.

The breaking of spherical symmetry, the formation and growth of various types of instabilities, and the turbulence within the bubble due to strong non-radial flows characterize the MS stage in 2D. These effects were missed by most earlier simulations.

The {subsequent} RSG wind has a much lower velocity than the MS wind, a higher mass-loss rate and a lower temperature. Therefore, it doesn't expand far into the MS wind, as seen in {the 1D simulations}, and piles up against the MS wind. It tries to form a new pressure equilibrium within the MS shell, although this may not always be possible. Hydrodynamically, the RSG wind will also be unstable, as shown in \citet{dwarkadas07}. Unfortunately this is not easily visible in Figure \ref{fig:den2d}c,d without zooming in on the wind region. An important aspect here is that the RSG wind is unable to {keep} the HII region {ionized, therefore it} begins to recombine, as seen in Figure \ref{fig:ion2d}b, which shows the ionization fraction of the wind bubble in the RSG phase. Depending on the length of the RSG phase, and the number of ionizing photons, the HII region may have enough time to become almost completely neutral. In this particular case, by the end of the RSG phase, the ionization fraction of the HII region has decreased to around 25--30\%. Since the wind does not travel far, the shape and structure of the hot bubble or the HII region are not significantly affected in this phase.

In the next phase, the fast W-R wind forms a mini wind bubble inside of the RSG wind. Its mass-loss rate being somewhat smaller, but its velocity significantly larger, its momentum tends to push the RSG shell outwards, breaking it up completely, and mixing the contents of both the RSG and W-R winds within the interior. As can be seen in Figure \ref{fig:den2d}e,f, the collision of the W-R wind with the RSG material leads to considerable fragmentation, and increases the turbulence within the bubble interior. A few large clumps and filaments can be seen to form within the hot bubble. These filaments are dense enough that they {may not be} fully ionized (Figure \ref{fig:ion2d}c). They are also cooler than the rest of the bubble. Some of the cold, dense filaments do not have a high enough temperature to produce X-rays, thus reducing the X-ray emissivity of the bubble. Simultaneously, the high surface temperature of the W-R star results in sufficient ionizing photons to re-ionize the HII region that had recombined significantly in the RSG stage. By the end of the simulation, the HII region is almost fully ionized. The W-R wind tends to push out significantly against the HII region, and in some simulations the wind makes its way all the way to the shell, with the ionized region all but disappearing. In this particular simulation, the hot wind region has occupied most of the ionized region (Figure \ref{fig:den2d}f), although in some parts the ionized region is still visible.

{The radius of the bubble in the 2D simulations is found to be smaller than that in the 1D simulations. While differences in grid resolution may contribute, this is mainly due to the fact that energy that would have gone into the expansion of the bubble is expended in other processes, thereby reducing the bubble expansion in multi-dimensions. These processes include: (1) The onset and growth of vorticity, and non-radial flow patterns, within the bubble. (2) Mixing of cold, dense clumps and filaments from the dense shell and ionized region into the hot, low density interior. The injection of mass into the shocked wind region leads to cooling of the hot shocked wind material. (3) Turbulent mixing due to various instabilities that arise at the interface of the hot shocked wind with the ionized region, and the ionized region with the dense shell. Coupled with ionization front instabilities, this leads to a high rate of cooling in some areas. Lancaster et al. \cite{lancasteretal21} and \mbox{\citet{km92}} have shown that efficient cooling can slow down the bubble expansion, as the bubble radius increases with time as t$^{0.5}$, slower than the t$^{0.6}$ solution given by Weaver et al. \cite{weaveretal77}. This will be explored in more detail in the paper describing the X-ray emission from the~bubble.}

\subsection{Ionization Fraction}

\begin{figure}[H]
\includegraphics[width=\linewidth]{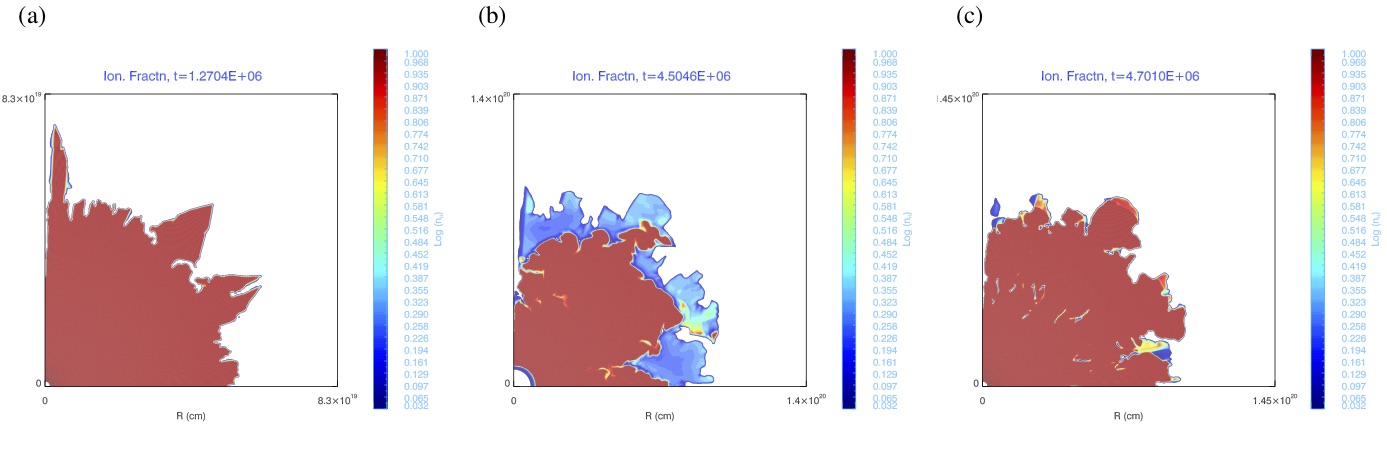}
\caption{{Evolution of the ionization fraction of the wind blown nebula over the various stages of the star's life. (\textbf{Left}) MS phase: the nebula is fully ionized. (\textbf{Middle}) RSG phase: the cooler star can no longer keep the HII region ionized, and it begins to recombine. (\textbf{Right}) W-R phase: the hot star re-ionizes almost the entire nebula.} }
\label{fig:ion2d}
\end{figure}

Figure \ref{fig:ion2d} shows the evolution of the ionization fraction in 2D. During the MS (\mbox{Figure \ref{fig:ion2d}a}) and W-R (Figure \ref{fig:ion2d}c) phases the wind bubble is almost fully ionized (red in the figure). In the RSG phase however (Figure \ref{fig:ion2d}b), as emphasized above, the star does not emit sufficient ionizing photons to keep the HII region ionized, and it begins to recombine.  In the subsequent W-R phase, the hot star has enough ionizing photons to ionize the entire nebula (Figure \ref{fig:ion2d}c). The distinction is of importance for any subsequent SN explosion; if the star were to end its life in the RSG phase (i.e., as a Type IIP or perhaps IIb SN), the resulting SN shock will evolve in a dense RSG wind, followed by a low density region and then a somewhat higher density but lower ionization region before colliding with the dense shell. Whereas if the star ends its life in the W-R phase, forming a Type Ib/c SN, the shock wave will evolve in a  generally low density, fully ionized medium before impacting the \mbox{dense shell.}

\section{Conclusions} \label{sec5}
We have described the evolution of the wind-blown bubble around a 40 $\msun$ star as it evolves over its lifetime, starting from the MS, veering off the MS to become a RSG star, and ending its life as a W-R star. We discussed the formation of both hydrodynamical and ionization front instabilities in the M-S phase, which have not been fully investigated in the literature in this specific context. The instabilities, {combined with turbulence within the interior,} result in the {formation of dense clumps and filaments, and} breaking of spherical symmetry within the bubble.  The ionization of the bubble through the various phases, and in particular the recombination of the outer ionized region in the RSG phase, was explored.  In a follow-up paper, the simulations will be used to compute the X-ray emission from \mbox{the bubble.}
\vspace{6pt}

\authorcontributions{}

\funding{VVD's research was funded by NSF grant 1911061, and by grants TM9-0001X and TM5-16001X, provided by NASA through the Chandra X-ray Observatory center, operated by SAO under NASA contract NAS8-03060.}

\institutionalreview{Not applicable 
}

\informedconsent{Not applicable 

}

\dataavailability{Figures from the simulations are shown in the text. Animations will be made available on the author's web site once the paper is published. We are happy to consider any requests from authors who may require the numerical output from the simulations; please email \mbox{the author.}}

\acknowledgments{We thank the reviewers for their thoughtful comments and suggestions, which have helped to improve the paper. The author is extremely grateful to Duane Rosenberg for providing the code that was modified and used to carry out the simulations described, and for his considerable help in setting it up and using it. We thank the organizers of APN8 for organizing a stimulating conference, and for providing the opportunity to publish this manuscript.}

\conflictsofinterest{The author declares no conflict of interest. The funders had no role in the design of the study; in the collection, analyses, or interpretation of data; in the writing of the manuscript, or in the decision to publish the results.} 
\begin{adjustwidth}{-\extralength}{0cm}

\reftitle{References}

\end{adjustwidth}
\end{document}